\documentclass[twocolumn,aps,prl,floatfix,superscriptaddress]{revtex4-1}
\usepackage{amsmath,amssymb}
\usepackage{graphicx,float}
\usepackage{times,txfonts}
\usepackage{textcomp}
\usepackage{color}
\usepackage{multirow}
\usepackage{color}
\usepackage{soul}
\usepackage{hyperref}
\usepackage{natbib}
\usepackage{nicefrac}
\usepackage{multirow}
\usepackage{blindtext}
\usepackage[export]{adjustbox}

\renewcommand{\epsilon}{\varepsilon}

\usepackage{sistyle}

\usepackage{soul}
\definecolor{hugoColor}{RGB}{59,134,255}
\definecolor{YellowOrange}{RGB}{226,154,2}
\sethlcolor{YellowOrange}
\sethlcolor{lightblue}

\sethlcolor{lightblue}

\begin{document}
\title{Generation of electron vortices using non-exact electric fields}

\author{Amir~H.~Tavabi}
\email{a.tavabi@fz-juelich.de}
\affiliation{Ernst Ruska-Centre for Microscopy and Spectroscopy with Electrons and Peter Gr\"unberg Institute, Forschungszentrum J\"ulich, 52425 J\"ulich, Germany}
\author{Hugo~Larocque}
\affiliation{Department of Physics, University of Ottawa, 25 Templeton St., Ottawa, Ontario, K1N 6N5 Canada}
\author{Peng-Han~Lu}
\affiliation{Ernst Ruska-Centre for Microscopy and Spectroscopy with Electrons and Peter Gr\"unberg Institute, Forschungszentrum J\"ulich, 52425 J\"ulich, Germany}
\author{Martial~Duchamp}
\affiliation{Ernst Ruska-Centre for Microscopy and Spectroscopy with Electrons and Peter Gr\"unberg Institute, Forschungszentrum J\"ulich, 52425 J\"ulich, Germany}
\affiliation{School of Materials Science and Engineering, Nanyang Technological University, 50 Nanyang Avenue, Singapore 639798, Singapore}
\author{Vincenzo~Grillo}
\affiliation{CNR-Istituto Nanoscienze, Centro S3, Via G Campi 213/a, I-41125 Modena, Italy}
\author{Ebrahim~Karimi}
\email{ekarimi@uottawa.ca}
\affiliation{Department of Physics, University of Ottawa, 25 Templeton St., Ottawa, Ontario, K1N 6N5 Canada}
\affiliation{Department of Physics, Institute for Advanced Studies in Basic Sciences, 45137-66731 Zanjan, Iran}
\author{Rafal~E.~Dunin-Borkowski}
\affiliation{Ernst Ruska-Centre for Microscopy and Spectroscopy with Electrons and Peter Gr\"unberg Institute, Forschungszentrum J\"ulich, 52425 J\"ulich, Germany}
\author{Giulio~Pozzi}
\affiliation{Ernst Ruska-Centre for Microscopy and Spectroscopy with Electrons and Peter Gr\"unberg Institute, Forschungszentrum J\"ulich, 52425 J\"ulich, Germany}
\affiliation{Department of Physics and Astronomy, University of Bologna, viale B. Pichat 6/2, 40127 Bologna, Italy}
%
\begin{abstract}
Vortices in electron beams can manifest several types of topological phenomena, such as the formation of exotic structures or interactions with topologically structured electromagnetic fields. For instance, the wavefunction of an electron beam can acquire a phase vortex upon propagating through a magnetic monopole, which, in practice, provides a convenient method for generating electron vortex beams. Here, we show how an electric field must be structured in order to achieve a similar effect. We find that, much as in the case of magnetic fields, closed but not exact electric fields can produce electron vortex beams. We proceed by fabricating a versatile near-obstruction-free device that is designed to approximately produce such fields and we systematically study their influence on incoming electron beams. With such a single device, electron vortex beams that are defined by a wide range of topological charges can be produced by means of a slight variation of an applied voltage. For this reason, this device is expected to be important in applications that rely on the sequential generation and manipulation of different types of electron vortices.

\end{abstract}
\pacs{Valid PACS appear here}
\maketitle

\noindent Vortices can generally be described as stagnant points surrounded by a form of coiling motion. These entities can, for instance, occur within complex fields, such as those describing the wavefunction of a quantum system or scalar optical waves. In such systems, vortices manifest themselves as singular points of the wavefield's phase, \emph{i.e.}, points around which the phase varies by an integer multiple $\ell$ of $2\pi$, where $\ell$ is referred to as the topological charge of the singularity~\cite{Bliokh:2017,Lloyd:2017_2,Larocque:2018}. In many cases, the presence of singularities in wavefields can lead to exotic forms of topological or geometric phenomena. For instance, the presence of a polarization singularity in a tightly focused optical wave can lead to the formation of a M\"obius strip~\cite{Bauer:2015}, while optical beams with carefully structured phase and polarization singularity distributions can result in the formation of knots~\cite{Dennis:2010,Larocque:2018aa}. Other types of singular behaviours can occur when matter waves of charged particles interact with structured electromagnetic fields. An example of such a phenomenon involves an electron beam propagating through a magnetic monopole. Upon propagating through the monopole's magnetic field, the electron experiences an azimuthally-dependent phase shift, thereby resulting in the formation of an electron vortex with a topological charge that is proportional to the strength of the monopole~\cite{Dirac:1948,Fukuhara:1983}. The impartment of this vortex arises directly from the electron's charge, in conjunction with the topological structure of the magnetic field. In spite of the physical elusiveness of magnetic monopoles, the above process can approximately occur in practice by replacing the monopole by the tip of a magnetic needle~\cite{Beche:2014,Blackburn:2014}, where the magnetic field closely resembles that required to impart a vortex onto the electron beam. When this vortex is located at the centre of the beam, the electron effectively acquires $\hbar \ell$ units of orbital angular momentum (OAM), where $\hbar$ is the reduced Planck constant~\cite{Bliokh:2007}. In conjunction with the electron's charge, the presence of a vortex also causes the electron to acquire a magnetic dipole moment $\ell \mu_B$, where $\mu_B$ is the Bohr magneton. This magnetic property makes OAM-carrying electrons desirable in materials science, as it allows them to be employed as nanoscale magnetic probes~\cite{Verbeeck:2010,Grillo:2017,Grillo:2017_2}. For this reason, magnetic needles offer an appealing alternative to other types of electron beam-shaping methods, such as diffractive holograms~\cite{Verbeeck:2010,Mcmorran:2011,Grillo:2014} and refractive phase masks ~\cite{Uchida:2010,Shiloh:2014}, which are both able to generate electron vortex beams. However, these devices have some technological constraints, including limited spatial resolution and an inability to generate electron vortices that are defined by arbitrary topological charges using a single device. For the specific case of magnetic needles, their magnetic nature prevents them from being positioned in the back focal plane of a magnetic lens and therefore to be used in electron imaging techniques that are analogous to optical spiral phase contrast microscopy~\cite{Jesacher:2005,Furhapter:2007}. Finally, the physical endurance of needles currently prevents them from generating electrons that are defined by larger values of OAM and are desirable in applications that rely on the stronger magnetic dipole moment that is carried by these electrons~\cite{Grillo:2017_2,Karlovets:2018}. In this \textit{Letter}, we discuss how an electric field can be shaped to achieve the same effect that a magnetic monopole has on an electron beam. The requirement to achieve such an effect relies deeply on the nature of the topology of electric fields, as compared to that of magnetic fields. We present an implementation of a device based on a recently proposed design~\cite{Pozzi:2017} that can be seen as an electric counterpart to a magnetic needle, and demonstrate how its electrical properties, in conjunction with its structural durability, can be used to generate electrons that carry a much broader range of tunable OAM values than those generated using magnetic needles.

\begin{figure}[t]
	\centering
	\includegraphics[width=\columnwidth]{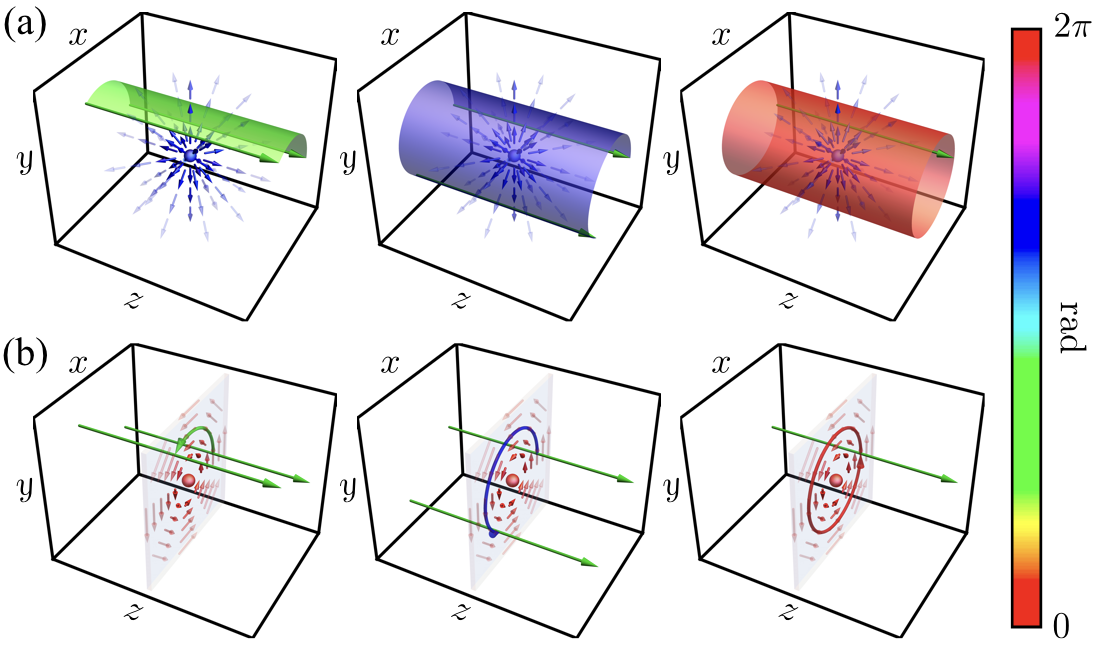}
	\caption{{\bf Propagation of electrons through closed but not exact electromagnetic fields}. (a) Surfaces used to calculate the relative azimuthal phase gained by electrons upon propagating through a magnetic monopole. (b) Lines used to calculate the relative azimuthal phase gained by electrons upon propagating through an azimuthally oriented electric field. Both the surfaces and the lines are coloured based on the phase obtained from integrating the fields over them. The green arrows in the plots represent the two trajectories that are used as boundaries to calculate their relative azimuthal phases. In these plots, field strengths that were designed to add a phase of $\theta=\varphi$ to the electron beam were considered.}
	\label{fig:fig1}
\end{figure}

In order to acquire OAM, electron waves need to acquire an azimuthally-dependent phase that results in the addition of an $\exp{(i\ell\varphi)}$ term to their mathematical formulation, where $\varphi$ is the transverse azimuthal coordinate. Such phases can be acquired by making the electron propagate through a potential whose action induces a structured phase shift~\cite{Mcmorran:2011,Grillo:2014,Uchida:2010,Shiloh:2014,Verbeeck:2018}. The phase $\theta$ acquired by an electron wave upon propagation through a system that is characterized by the presence of an electromagnetic field can be expressed in the form~\cite{Bliokh:2007}
\begin{equation}
	\label{eq:electronPhase}
	\theta = \underbrace{\frac{1}{\hbar} \int (\mathbf{p} \cdot d\mathbf{r}-\mathcal{E}\,dt)}_{\text{Dynamic Phase}} +  \underbrace{\frac{e}{\hbar} \int \mathbf{A} \cdot d\mathbf{r}}_{\text{Dirac Phase}} +  \underbrace{\int \boldsymbol{\mathcal{A}} \cdot d\mathbf{R}}_{\text{Berry Phase}}.
\end{equation}
The first term in this equation accounts for the dynamical phase acquired by the electron, where $\mathbf{p}$ is the electron's kinetic momentum, $\mathcal{E}=p^2/2m+e\Phi-\boldsymbol{\mu}\cdot\mathbf{B}$ is its energy, $m$ is its mass, $e$ is its charge, $\Phi$ is the field's scalar potential, $\boldsymbol{\mu}$ is the electron's magnetic dipole moment, and $\mathbf{B}$ is the magnetic field. This phase is typically used in the holographic generation of OAM-carrying electrons, given that such methods rely on devices, which have a mean inner potential that affects both the energy and the momentum of propagating electrons. The second term is the Dirac phase, where $\mathbf{A}$ is the field's vector potential. It is responsible for the impartment of OAM onto electrons propagating through a magnetic monopole~\cite{Fukuhara:1983} and other phenomena such as the Aharonov-Bohm effect~\cite{Aharonov:1959}. The last term represents the Berry phase, where $\boldsymbol{\mathcal{A}}$ is the Berry curvature associated with an adiabatically varied parameter $\mathbf{R}$.

The OAM acquired by an electron beam exposed to a magnetic monopole can be found by calculating the relative Dirac phase attributed to propagation along different transverse azimuthal angles. Such a calculation can be performed in a cylindrical coordinate system $(\rho,\varphi,z)$, where the $z$-axis is set to lie along the electron's direction of propagation. The phase acquired by electrons propagating along $\varphi=0$ is set to zero, allowing it to be taken as a reference, with respect to which the phase of electrons travelling along other values of $\varphi$ can be calculated. As shown in Fig.~\ref{fig:fig1}(a), this approach effectively enables the use of Stokes' theorem to calculate the acquired Dirac phase, \emph{i.e.}, $\int {A}_z\,d{z} = \int \mathbf{B} \cdot\,\rho d\varphi \wedge dz$, where $\mathbf{B}=(\mu_0 q_m)/(4\pi r^3)\,\mathbf{r}$ is the magnetic field attributed to the monopole, $\mu_0$ is the permeability of free-space, $q_m$ is the magnetic charge of the monopole, $\mathbf{r}$ is the spherical radial coordinate, and $\wedge$ denotes the exterior product. Integrating the above expression reveals that the electrons acquire an azimuthal phase given by $\theta=(e \mu_0 q_m / h) \varphi$, implying that the impartment of discrete units of OAM requires a monopole strength $q_m$ set to an integer multiple of $h/(e \mu_0)$.

Some remarks concerning the above analysis can be made with reference to the topology of the monopole's magnetic field and how it relates to the OAM acquired by the electron. The magnetic field formally consists of a differential form, more specifically a 2-form, which is defined over the manifold $\mathbb{R}^3-\{\mathbf{0}\}$, where the exclusion of the origin results from the spherical radial coordinate not being defined at this point. This exclusion results in the formation of a so-called ``2-hole'' at the origin. An arbitrary two-dimensional surface in the manifold cannot be deformed into another without crossing the hole, causing the magnetic field to be closed but not exact. This non-simply-connected space allows a propagating electron to acquire OAM. The radial nature of the magnetic field enables an azimuthally increasing flux through the surfaces shown in Fig.~\ref{fig:fig1}(a), and can be used to calculate the azimuthal phase gained by the beam.

Unlike magnetic fields, the differential forms that describe electric fields consist of 1-forms, implying that their topology must be different to be used to impart OAM to incoming electrons. In order to achieve such an effect, the electric field can also be expected to be closed yet not exact due to the presence of a ``1-hole'' in two-dimensional planes that are perpendicular to the propagation of the electron beam. Furthermore, the phase acquired due to the presence of the electric field can also be expected to rely on a line integral, \emph{i.e.}, a 1+1 dimensional boundary, as opposed to the surface used in the case of the magnetic field, which consists of a 2+1 dimensional boundary. This intuitive approach can be concretized by examining the contribution of the electric field to the phase acquired by an electron due to the presence of an electric field in Eq.~(\ref{eq:electronPhase}), \emph{i.e.}, $- (e/\hbar) \int \Phi \,dt$~\cite{Boyer:1973}. Assuming the presence of a static electric field along with a paraxial configuration over which the electrons are propagating along the $z$-axis, the integration over time can be replaced by an integration along $z$. As a result, the acquired phase becomes $-(e m/\hbar p_0) \int \Phi \,dz$, implying that the potential $\Phi$ needs to be monotonic along $\varphi$ over a certain range $\Delta z$ to impart OAM onto electrons. Such a requirement can also be seen by expressing the azimuthal domain of this potential with respect to the azimuthal component of the electric field $E_\varphi$, i.e. $\Phi(\varphi,z) = -\int E_\varphi(\varphi,z) \,\rho d\varphi$. In order to impart a phase that increases linearly with azimuthal angle, an electric field with a constant azimuthal component is required. This requirement is shown schematically in Fig.~\ref{fig:fig1}(b). As initially postulated, the azimuthal nature of the electric field makes it not exact and constrains it to the manifold $\mathbb{R}^2-\{\mathbf{0}\}$ for a given $z$ value, requiring the presence of a ``1-hole'' at the origin. Moreover, the presence of the azimuthally varying phase ultimately relies on a line integral as opposed to a surface integral.
\begin{figure}[t]
	\centering
	\includegraphics[width=\linewidth]{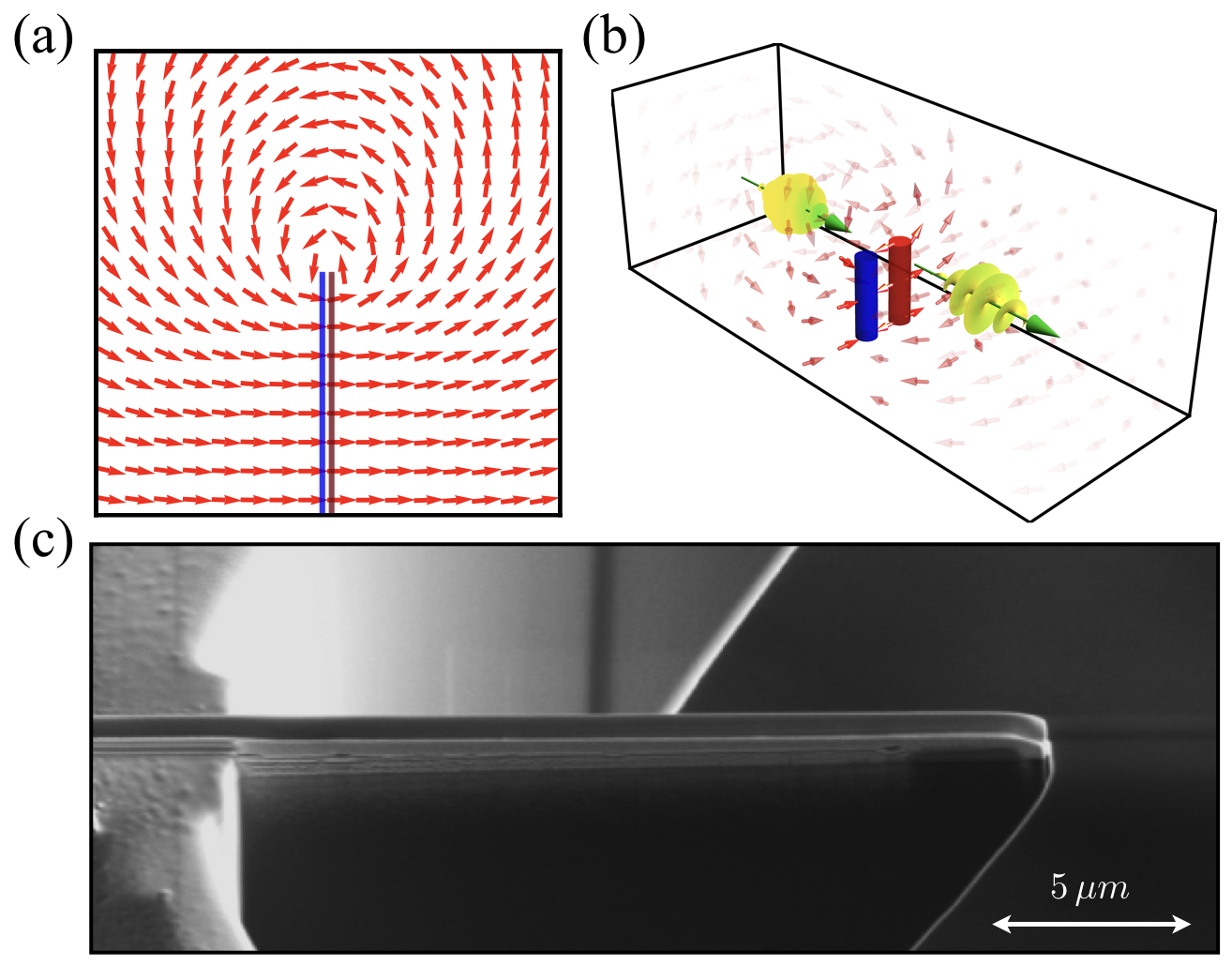}
	\caption{{\bf Device generating a tunable, closed, but not exact electric field}. Two sub-micron wires in close proximity with different electric potentials can effectively produce an azimuthally oriented electric field at their extremity. (a) Electric field resulting from such a configuration at the tips of both wires, which, represented as red and blue segments, are infinitely long and infinitely close to one another. (b) Device configuration required for the impartment of OAM onto an electron beam by means of such an electric field. The electric field is represented by red arrows, with a relative opacity that is dictated by its relative strength. (c) Scanning electron micrograph of a fabricated device, which consists of two 200-nm-wide and 15-\textmu m-long nanowires separated by a 200~nm gap, which lie on a silicon-nitride/silicon substrate.}
	\label{fig:fig2}
\end{figure}
In order to produce such an azimuthal electric field, one can adopt an approach similar to that for the extremity of a magnetic needle to approximately replicate the magnetic field of a monopole. In essence, the approach consists of using a dipole-like structure that locally displays fields with the geometry required to impart OAM onto electrons~\cite{Pozzi:2017}. These local features can thereafter be enhanced by modifying the local geometry of the structure itself. As illustrated in Fig.~\ref{fig:fig2}(a), the tip of an electric dipole structure consisting of two elongated and extremely close charged rods allows for the generation of an almost perfectly azimuthal electric field. By varying the charges on the two rods, the relative strength of the azimuthal electric field can be modified. Schematic diagrams of how such a device can be used to impart OAM onto incoming electrons are shown in Fig.~\ref{fig:fig2}(b). By varying the relative potential between the two rods, or equivalently the effective charge that they carry, the strength of the azimuthal electric field can be adjusted, thereby enabling the generation of electron vortices that are defined by a tunable amount of OAM. Because the strength of the electric field varies continuously as a function of relative potential, only discrete values of voltage applied between the two rods can lead to quantized azimuthal phase variations attributed to OAM. 

In order to fabricate such a device, we adopted a fabrication procedure that involved the combined use of electron beam lithography and focused ion beam (FIB) milling. This approach enabled the fabrication of two 200-nm-wide and 15-\textmu m-long metallic wires separated by a 200~nm gap, which were patterned lithographically onto a silicon nitride/silicon substrate. A semi-circular opening with a radius of 15 \textmu m around the wires was then created using FIB milling. This design enables the substrate to be grounded to the microscope, while the wires are connected to an external voltage source, thereby preventing the formation of a short circuit between the wires. Given that the wires only span a bridge of 600~nm over the 30 \textmu m circular opening, the device is almost obstruction-free, allowing higher transmission efficiency and reducing potential artefacts from scattering introduced by material-based phase masks. A scanning electron micrograph of the device is shown in Fig.~\ref{fig:fig2}(c). A device achieving a similar effect, consisting of a dielectric rod with a partial metallic coating, was previously reported~\cite{Blackburn:2016}. However, details about how the topology of the electric field explicitly interacts with that of the electron beam were not explored. Moreover, unlike for the design reported in the present work, the latter device was not tunable, thereby preventing more extensive studies involving the role of the strength and the orientation of the azimuthal electric field on the generated electron vortex.\newline
\begin{figure*}[t]
	\centering
	\includegraphics[width=\linewidth]{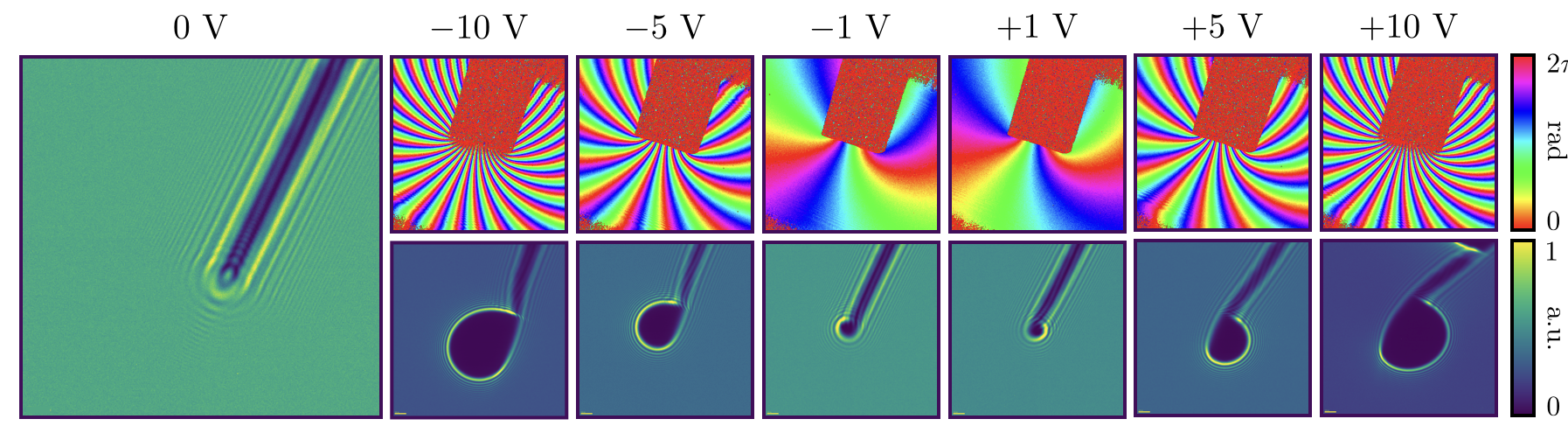}
	\caption[]{{\bf Transverse spatial profile of electrons affected by the device}. Left: TEM image of the device while the two wires are held at the same potential obtained from Fresnel imaging with a nominal defocus near 25 mm. Top row: acquired phase of an electron wavefunction having interacted with the device while the two wires were held at various potentials. These phase profiles were reconstructed by means of off-axis electron holography defined by a fringe spacing of 1.9 nm. Both the absolute value and the sign of the generated electron vortex beam's topological charge are observed to follow that of the applied voltage. Bottom row: corresponding probability density functions of the electrons obtained from TEM images by means of Fresnel imaging experiments.}
	\label{fig:fig3}
\end{figure*}

The device's ability to impart a vortex on an electron was demonstrated by measuring it inside a transmission electron microscope (TEM, FEI Titan 60-300) equipped with a Schottky-type high brightness field emission gun (FEI X-FEG) and two electron biprisms. The microscope was operated at 300 kV during the measurements. An electron biprism was used to form an interference region with a 1.5-\textmu m-wide field of view. A 1.9 nm holographic interference fringe spacing was obtained by using a biprism voltage at 107~V. Off-axis electron holography measurements obtained in this configuration were used to reconstruct the phase profile of electrons that had interacted with the device. Representative phase profiles are shown in the top row of Fig.~\ref{fig:fig3} and clearly display azimuthal variations attributed to the presence of a phase vortex.

As expected, the strength of these variations increases with that of the device's azimuthal electric field, which is tuned by means of applying a voltage difference between the two wires, thereby enabling the generation of electron vortices that are defined by topological charges of up to $\pm 30$. Furthermore, the handedness of the electron beam's azimuthal phase profile is observed to depend on the sign of the electric field, which is controlled by switching the sign of the applied voltage.
The presence of the vortices is also attested by the profile of the probability density function of the electrons' wavefunction. Defocused images recorded in the Fresnel domain provide a means of examining the latter quantity, given that they are acquired over many electrons and that the use of a larger defocus amplifies the small deflection of the electrons, thereby yielding a good reflection of the probabilistic nature of their wavefunction. The images that were obtained are shown in the bottom row of Fig.~\ref{fig:fig3}, for a series of different potentials applied to the two wires, as well as on the far-left side of this figure, where no potential was applied. They display the expected presence of a null in the electron's probability density function located at the position of the phase vortex. The transverse extent of these nulls also displays the quintessential trend of increasing with the absolute value of the beam's topological charge.
The phase and probability density profiles of the electron vortices are in good agreement with numerical results reported in an earlier study~\cite{Pozzi:2017}. Slight discrepancies between the simulations and experiments arise primarily from minor details regarding how the device is fabricated. As the silicon nitride/silicon support of the wires becomes charged when it is being used, the presence of the support introduces additional contributions to the potential, akin to that of a biprism~\cite{Pozzi:2017}. The finite length of the wires also contributes to discrepancies in the potential. These limitations can, however, be addressed with slight design modifications, such as the addition of static fields, as well as by ensuring that the wires can be biased independently.

In summary, we have demonstrated the tunable generation of electron vortex beams by means of closed but not exact electric fields. On a fundamental level, this generation scheme holds several similarities to the use of magnetic monopoles, which have closed but not exact magnetic fields, as a means to achieve a similar effect. However, the differences between the differential forms that describe electric and magnetic fields manifest themselves within the structures of the fields that can be used to produce electron vortices. We were able to construct such fields by means of a nanofabricated device consisting of two wires held at different potentials. By adjusting the potential difference between the wires, we were able to adjust both the strength and the sign of the device's electric field, thereby enabling the generation of electron vortex beams defined by arbitrary topological charges. Besides demonstrating the relationship between the differential geometry of an electric field and the vortex content of an electron affected by it, our device holds significant promise for applications that rely on the sequential generation of electron vortex beams defined by different topological charges. Potential extensions of this device could include the use of larger numbers of electrodes to generate more complicated electron beam shapes -- perhaps three-dimensional ones, especially if the electric fields can be modulated on a timescale over which the electron wave passes the device.

\noindent The authors acknowledge the support of the European Union's Horizon 2020 Research and Innovation Programme (Q-SORT), grant number 766970. P.L. and R.E.D.B. acknowledge the support of the Deutsche Forschungsgemeinschaft for a Deutsch-Israelische Projektkooperation (DIP) Grant. R.E.D.B. is grateful for funding from the European Research Council under the European Union's Seventh Framework Programme (FP7/2007-2013)/ERC grant agreement number 320832. H.L. and E.K. acknowledge the support of Canada Research Chairs (CRC), and Ontario's Early Researcher Award (ERA). M.D. acknowledges the financial support of the Ministry of Education Academic Research Fund Tier 1 grants RG101/17. The authors thank Maximilian Kruth for his help with the FIB work.


\begin{thebibliography}{26}%
\makeatletter
\providecommand \@ifxundefined [1]{%
 \@ifx{#1\undefined}
}%
\providecommand \@ifnum [1]{%
 \ifnum #1\expandafter \@firstoftwo
 \else \expandafter \@secondoftwo
 \fi
}%
\providecommand \@ifx [1]{%
 \ifx #1\expandafter \@firstoftwo
 \else \expandafter \@secondoftwo
 \fi
}%
\providecommand \natexlab [1]{#1}%
\providecommand \enquote  [1]{``#1''}%
\providecommand \bibnamefont  [1]{#1}%
\providecommand \bibfnamefont [1]{#1}%
\providecommand \citenamefont [1]{#1}%
\providecommand \href@noop [0]{\@secondoftwo}%
\providecommand \href [0]{\begingroup \@sanitize@url \@href}%
\providecommand \@href[1]{\@@startlink{#1}\@@href}%
\providecommand \@@href[1]{\endgroup#1\@@endlink}%
\providecommand \@sanitize@url [0]{\catcode `\\12\catcode `\$12\catcode
  `\&12\catcode `\#12\catcode `\^12\catcode `\_12\catcode `\%12\relax}%
\providecommand \@@startlink[1]{}%
\providecommand \@@endlink[0]{}%
\providecommand \url  [0]{\begingroup\@sanitize@url \@url }%
\providecommand \@url [1]{\endgroup\@href {#1}{\urlprefix }}%
\providecommand \urlprefix  [0]{URL }%
\providecommand \Eprint [0]{\href }%
\providecommand \doibase [0]{http://dx.doi.org/}%
\providecommand \selectlanguage [0]{\@gobble}%
\providecommand \bibinfo  [0]{\@secondoftwo}%
\providecommand \bibfield  [0]{\@secondoftwo}%
\providecommand \translation [1]{[#1]}%
\providecommand \BibitemOpen [0]{}%
\providecommand \bibitemStop [0]{}%
\providecommand \bibitemNoStop [0]{.\EOS\space}%
\providecommand \EOS [0]{\spacefactor3000\relax}%
\providecommand \BibitemShut  [1]{\csname bibitem#1\endcsname}%
\let\auto@bib@innerbib\@empty
\bibitem [{\citenamefont {Bliokh}\ \emph {et~al.}(2017)\citenamefont {Bliokh},
  \citenamefont {Ivanov}, \citenamefont {Guzzinati}, \citenamefont {Clark},
  \citenamefont {Van~Boxem}, \citenamefont {B{\'e}ch{\'e}}, \citenamefont
  {Juchtmans}, \citenamefont {Alonso}, \citenamefont {Schattschneider},
  \citenamefont {Nori} \emph {et~al.}}]{Bliokh:2017}%
  \BibitemOpen
  \bibfield  {author} {\bibinfo {author} {\bibfnamefont {K.}~\bibnamefont
  {Bliokh}}, \bibinfo {author} {\bibfnamefont {I.}~\bibnamefont {Ivanov}},
  \bibinfo {author} {\bibfnamefont {G.}~\bibnamefont {Guzzinati}}, \bibinfo
  {author} {\bibfnamefont {L.}~\bibnamefont {Clark}}, \bibinfo {author}
  {\bibfnamefont {R.}~\bibnamefont {Van~Boxem}}, \bibinfo {author}
  {\bibfnamefont {A.}~\bibnamefont {B{\'e}ch{\'e}}}, \bibinfo {author}
  {\bibfnamefont {R.}~\bibnamefont {Juchtmans}}, \bibinfo {author}
  {\bibfnamefont {M.}~\bibnamefont {Alonso}}, \bibinfo {author} {\bibfnamefont
  {P.}~\bibnamefont {Schattschneider}}, \bibinfo {author} {\bibfnamefont
  {F.}~\bibnamefont {Nori}},  \emph {et~al.},\ }\href@noop {} {\bibfield
  {journal} {\bibinfo  {journal} {Physics Reports}\ }\textbf {\bibinfo {volume}
  {690}},\ \bibinfo {pages} {1} (\bibinfo {year} {2017})}\BibitemShut {NoStop}%
\bibitem [{\citenamefont {Lloyd}\ \emph {et~al.}(2017)\citenamefont {Lloyd},
  \citenamefont {Babiker}, \citenamefont {Thirunavukkarasu},\ and\
  \citenamefont {Yuan}}]{Lloyd:2017_2}%
  \BibitemOpen
  \bibfield  {author} {\bibinfo {author} {\bibfnamefont {S.}~\bibnamefont
  {Lloyd}}, \bibinfo {author} {\bibfnamefont {M.}~\bibnamefont {Babiker}},
  \bibinfo {author} {\bibfnamefont {G.}~\bibnamefont {Thirunavukkarasu}}, \
  and\ \bibinfo {author} {\bibfnamefont {J.}~\bibnamefont {Yuan}},\ }\href@noop
  {} {\bibfield  {journal} {\bibinfo  {journal} {Reviews of Modern Physics}\
  }\textbf {\bibinfo {volume} {89}},\ \bibinfo {pages} {035004} (\bibinfo
  {year} {2017})}\BibitemShut {NoStop}%
\bibitem [{\citenamefont {Larocque}\ \emph
  {et~al.}(2018{\natexlab{a}})\citenamefont {Larocque}, \citenamefont
  {Kaminer}, \citenamefont {Grillo}, \citenamefont {Leuchs}, \citenamefont
  {Padgett}, \citenamefont {Boyd}, \citenamefont {Segev},\ and\ \citenamefont
  {Karimi}}]{Larocque:2018}%
  \BibitemOpen
  \bibfield  {author} {\bibinfo {author} {\bibfnamefont {H.}~\bibnamefont
  {Larocque}}, \bibinfo {author} {\bibfnamefont {I.}~\bibnamefont {Kaminer}},
  \bibinfo {author} {\bibfnamefont {V.}~\bibnamefont {Grillo}}, \bibinfo
  {author} {\bibfnamefont {G.}~\bibnamefont {Leuchs}}, \bibinfo {author}
  {\bibfnamefont {M.~J.}\ \bibnamefont {Padgett}}, \bibinfo {author}
  {\bibfnamefont {R.~W.}\ \bibnamefont {Boyd}}, \bibinfo {author}
  {\bibfnamefont {M.}~\bibnamefont {Segev}}, \ and\ \bibinfo {author}
  {\bibfnamefont {E.}~\bibnamefont {Karimi}},\ }\href@noop {} {\bibfield
  {journal} {\bibinfo  {journal} {Contemporary Physics}\ }\textbf {\bibinfo
  {volume} {59}},\ \bibinfo {pages} {126} (\bibinfo {year}
  {2018}{\natexlab{a}})}\BibitemShut {NoStop}%
\bibitem [{\citenamefont {Bauer}\ \emph {et~al.}(2015)\citenamefont {Bauer},
  \citenamefont {Banzer}, \citenamefont {Karimi}, \citenamefont {Orlov},
  \citenamefont {Rubano}, \citenamefont {Marrucci}, \citenamefont {Santamato},
  \citenamefont {Boyd},\ and\ \citenamefont {Leuchs}}]{Bauer:2015}%
  \BibitemOpen
  \bibfield  {author} {\bibinfo {author} {\bibfnamefont {T.}~\bibnamefont
  {Bauer}}, \bibinfo {author} {\bibfnamefont {P.}~\bibnamefont {Banzer}},
  \bibinfo {author} {\bibfnamefont {E.}~\bibnamefont {Karimi}}, \bibinfo
  {author} {\bibfnamefont {S.}~\bibnamefont {Orlov}}, \bibinfo {author}
  {\bibfnamefont {A.}~\bibnamefont {Rubano}}, \bibinfo {author} {\bibfnamefont
  {L.}~\bibnamefont {Marrucci}}, \bibinfo {author} {\bibfnamefont
  {E.}~\bibnamefont {Santamato}}, \bibinfo {author} {\bibfnamefont {R.~W.}\
  \bibnamefont {Boyd}}, \ and\ \bibinfo {author} {\bibfnamefont
  {G.}~\bibnamefont {Leuchs}},\ }\href@noop {} {\bibfield  {journal} {\bibinfo
  {journal} {Science}\ }\textbf {\bibinfo {volume} {347}},\ \bibinfo {pages}
  {964} (\bibinfo {year} {2015})}\BibitemShut {NoStop}%
\bibitem [{\citenamefont {Dennis}\ \emph {et~al.}(2010)\citenamefont {Dennis},
  \citenamefont {King}, \citenamefont {Jack}, \citenamefont {O'Holleran},\ and\
  \citenamefont {Padgett}}]{Dennis:2010}%
  \BibitemOpen
  \bibfield  {author} {\bibinfo {author} {\bibfnamefont {M.~R.}\ \bibnamefont
  {Dennis}}, \bibinfo {author} {\bibfnamefont {R.~P.}\ \bibnamefont {King}},
  \bibinfo {author} {\bibfnamefont {B.}~\bibnamefont {Jack}}, \bibinfo {author}
  {\bibfnamefont {K.}~\bibnamefont {O'Holleran}}, \ and\ \bibinfo {author}
  {\bibfnamefont {M.~J.}\ \bibnamefont {Padgett}},\ }\href@noop {} {\bibfield
  {journal} {\bibinfo  {journal} {Nature Physics}\ }\textbf {\bibinfo {volume}
  {6}},\ \bibinfo {pages} {118} (\bibinfo {year} {2010})}\BibitemShut {NoStop}%
\bibitem [{\citenamefont {Larocque}\ \emph
  {et~al.}(2018{\natexlab{b}})\citenamefont {Larocque}, \citenamefont {Sugic},
  \citenamefont {Mortimer}, \citenamefont {Taylor}, \citenamefont {Fickler},
  \citenamefont {Boyd}, \citenamefont {Dennis},\ and\ \citenamefont
  {Karimi}}]{Larocque:2018aa}%
  \BibitemOpen
  \bibfield  {author} {\bibinfo {author} {\bibfnamefont {H.}~\bibnamefont
  {Larocque}}, \bibinfo {author} {\bibfnamefont {D.}~\bibnamefont {Sugic}},
  \bibinfo {author} {\bibfnamefont {D.}~\bibnamefont {Mortimer}}, \bibinfo
  {author} {\bibfnamefont {A.~J.}\ \bibnamefont {Taylor}}, \bibinfo {author}
  {\bibfnamefont {R.}~\bibnamefont {Fickler}}, \bibinfo {author} {\bibfnamefont
  {R.~W.}\ \bibnamefont {Boyd}}, \bibinfo {author} {\bibfnamefont {M.~R.}\
  \bibnamefont {Dennis}}, \ and\ \bibinfo {author} {\bibfnamefont
  {E.}~\bibnamefont {Karimi}},\ }\href {\doibase 10.1038/s41567-018-0229-2}
  {\bibfield  {journal} {\bibinfo  {journal} {Nature Physics}\ } (\bibinfo
  {year} {2018}{\natexlab{b}}),\ 10.1038/s41567-018-0229-2}\BibitemShut
  {NoStop}%
\bibitem [{\citenamefont {Dirac}(1948)}]{Dirac:1948}%
  \BibitemOpen
  \bibfield  {author} {\bibinfo {author} {\bibfnamefont {P.~A.~M.}\
  \bibnamefont {Dirac}},\ }\href@noop {} {\bibfield  {journal} {\bibinfo
  {journal} {Physical Review}\ }\textbf {\bibinfo {volume} {74}},\ \bibinfo
  {pages} {817} (\bibinfo {year} {1948})}\BibitemShut {NoStop}%
\bibitem [{\citenamefont {Fukuhara}\ \emph {et~al.}(1983)\citenamefont
  {Fukuhara}, \citenamefont {Shinagawa}, \citenamefont {Tonomura},\ and\
  \citenamefont {Fujiwara}}]{Fukuhara:1983}%
  \BibitemOpen
  \bibfield  {author} {\bibinfo {author} {\bibfnamefont {A.}~\bibnamefont
  {Fukuhara}}, \bibinfo {author} {\bibfnamefont {K.}~\bibnamefont {Shinagawa}},
  \bibinfo {author} {\bibfnamefont {A.}~\bibnamefont {Tonomura}}, \ and\
  \bibinfo {author} {\bibfnamefont {H.}~\bibnamefont {Fujiwara}},\ }\href@noop
  {} {\bibfield  {journal} {\bibinfo  {journal} {Physical Review B}\ }\textbf
  {\bibinfo {volume} {27}},\ \bibinfo {pages} {1839} (\bibinfo {year}
  {1983})}\BibitemShut {NoStop}%
\bibitem [{\citenamefont {B{\'e}ch{\'e}}\ \emph {et~al.}(2014)\citenamefont
  {B{\'e}ch{\'e}}, \citenamefont {Van~Boxem}, \citenamefont {Van~Tendeloo},\
  and\ \citenamefont {Verbeeck}}]{Beche:2014}%
  \BibitemOpen
  \bibfield  {author} {\bibinfo {author} {\bibfnamefont {A.}~\bibnamefont
  {B{\'e}ch{\'e}}}, \bibinfo {author} {\bibfnamefont {R.}~\bibnamefont
  {Van~Boxem}}, \bibinfo {author} {\bibfnamefont {G.}~\bibnamefont
  {Van~Tendeloo}}, \ and\ \bibinfo {author} {\bibfnamefont {J.}~\bibnamefont
  {Verbeeck}},\ }\href@noop {} {\bibfield  {journal} {\bibinfo  {journal}
  {Nature Physics}\ }\textbf {\bibinfo {volume} {10}},\ \bibinfo {pages} {26}
  (\bibinfo {year} {2014})}\BibitemShut {NoStop}%
\bibitem [{\citenamefont {Blackburn}\ and\ \citenamefont
  {Loudon}(2014)}]{Blackburn:2014}%
  \BibitemOpen
  \bibfield  {author} {\bibinfo {author} {\bibfnamefont {A.~M.}\ \bibnamefont
  {Blackburn}}\ and\ \bibinfo {author} {\bibfnamefont {J.~C.}\ \bibnamefont
  {Loudon}},\ }\href {\doibase
  http://dx.doi.org/10.1016/j.ultramic.2013.08.009} {\bibfield  {journal}
  {\bibinfo  {journal} {Ultramicroscopy}\ }\textbf {\bibinfo {volume} {136}},\
  \bibinfo {pages} {127} (\bibinfo {year} {2014})}\BibitemShut {NoStop}%
\bibitem [{\citenamefont {Bliokh}\ \emph {et~al.}(2007)\citenamefont {Bliokh},
  \citenamefont {Bliokh}, \citenamefont {Savel'Ev},\ and\ \citenamefont
  {Nori}}]{Bliokh:2007}%
  \BibitemOpen
  \bibfield  {author} {\bibinfo {author} {\bibfnamefont {K.~Y.}\ \bibnamefont
  {Bliokh}}, \bibinfo {author} {\bibfnamefont {Y.~P.}\ \bibnamefont {Bliokh}},
  \bibinfo {author} {\bibfnamefont {S.}~\bibnamefont {Savel'Ev}}, \ and\
  \bibinfo {author} {\bibfnamefont {F.}~\bibnamefont {Nori}},\ }\href@noop {}
  {\bibfield  {journal} {\bibinfo  {journal} {Physical Review Letters}\
  }\textbf {\bibinfo {volume} {99}},\ \bibinfo {pages} {190404} (\bibinfo
  {year} {2007})}\BibitemShut {NoStop}%
\bibitem [{\citenamefont {Verbeeck}\ \emph {et~al.}(2010)\citenamefont
  {Verbeeck}, \citenamefont {Tian},\ and\ \citenamefont
  {Schattschneider}}]{Verbeeck:2010}%
  \BibitemOpen
  \bibfield  {author} {\bibinfo {author} {\bibfnamefont {J.}~\bibnamefont
  {Verbeeck}}, \bibinfo {author} {\bibfnamefont {H.}~\bibnamefont {Tian}}, \
  and\ \bibinfo {author} {\bibfnamefont {P.}~\bibnamefont {Schattschneider}},\
  }\href@noop {} {\bibfield  {journal} {\bibinfo  {journal} {Nature}\ }\textbf
  {\bibinfo {volume} {467}},\ \bibinfo {pages} {301} (\bibinfo {year}
  {2010})}\BibitemShut {NoStop}%
\bibitem [{\citenamefont {Grillo}\ \emph
  {et~al.}(2017{\natexlab{a}})\citenamefont {Grillo}, \citenamefont {Tavabi},
  \citenamefont {Venturi}, \citenamefont {Larocque}, \citenamefont {Balboni},
  \citenamefont {Gazzadi}, \citenamefont {Frabboni}, \citenamefont {Lu},
  \citenamefont {Mafakheri}, \citenamefont {Bouchard}, \citenamefont
  {Dunin-Borkowski}, \citenamefont {Boyd}, \citenamefont {Lavery},
  \citenamefont {Padgett},\ and\ \citenamefont {Karimi}}]{Grillo:2017}%
  \BibitemOpen
  \bibfield  {author} {\bibinfo {author} {\bibfnamefont {V.}~\bibnamefont
  {Grillo}}, \bibinfo {author} {\bibfnamefont {A.~H.}\ \bibnamefont {Tavabi}},
  \bibinfo {author} {\bibfnamefont {F.}~\bibnamefont {Venturi}}, \bibinfo
  {author} {\bibfnamefont {H.}~\bibnamefont {Larocque}}, \bibinfo {author}
  {\bibfnamefont {R.}~\bibnamefont {Balboni}}, \bibinfo {author} {\bibfnamefont
  {G.~C.}\ \bibnamefont {Gazzadi}}, \bibinfo {author} {\bibfnamefont
  {S.}~\bibnamefont {Frabboni}}, \bibinfo {author} {\bibfnamefont {P.-H.}\
  \bibnamefont {Lu}}, \bibinfo {author} {\bibfnamefont {E.}~\bibnamefont
  {Mafakheri}}, \bibinfo {author} {\bibfnamefont {F.}~\bibnamefont {Bouchard}},
  \bibinfo {author} {\bibfnamefont {R.~E.}\ \bibnamefont {Dunin-Borkowski}},
  \bibinfo {author} {\bibfnamefont {R.~W.}\ \bibnamefont {Boyd}}, \bibinfo
  {author} {\bibfnamefont {M.~P.~J.}\ \bibnamefont {Lavery}}, \bibinfo {author}
  {\bibfnamefont {M.~J.}\ \bibnamefont {Padgett}}, \ and\ \bibinfo {author}
  {\bibfnamefont {E.}~\bibnamefont {Karimi}},\ }\href@noop {} {\bibfield
  {journal} {\bibinfo  {journal} {Nature Communications}\ }\textbf {\bibinfo
  {volume} {8}},\ \bibinfo {pages} {15536} (\bibinfo {year}
  {2017}{\natexlab{a}})}\BibitemShut {NoStop}%
\bibitem [{\citenamefont {Grillo}\ \emph
  {et~al.}(2017{\natexlab{b}})\citenamefont {Grillo}, \citenamefont {Harvey},
  \citenamefont {Venturi}, \citenamefont {Pierce}, \citenamefont {Balboni},
  \citenamefont {Bouchard}, \citenamefont {Gazzadi}, \citenamefont {Frabboni},
  \citenamefont {Tavabi}, \citenamefont {Li}, \citenamefont {Dunin-Borkowski},
  \citenamefont {Boyd}, \citenamefont {McMorran},\ and\ \citenamefont
  {Karimi}}]{Grillo:2017_2}%
  \BibitemOpen
  \bibfield  {author} {\bibinfo {author} {\bibfnamefont {V.}~\bibnamefont
  {Grillo}}, \bibinfo {author} {\bibfnamefont {T.~R.}\ \bibnamefont {Harvey}},
  \bibinfo {author} {\bibfnamefont {F.}~\bibnamefont {Venturi}}, \bibinfo
  {author} {\bibfnamefont {J.~S.}\ \bibnamefont {Pierce}}, \bibinfo {author}
  {\bibfnamefont {R.}~\bibnamefont {Balboni}}, \bibinfo {author} {\bibfnamefont
  {F.}~\bibnamefont {Bouchard}}, \bibinfo {author} {\bibfnamefont {G.~C.}\
  \bibnamefont {Gazzadi}}, \bibinfo {author} {\bibfnamefont {S.}~\bibnamefont
  {Frabboni}}, \bibinfo {author} {\bibfnamefont {A.~H.}\ \bibnamefont
  {Tavabi}}, \bibinfo {author} {\bibfnamefont {Z.-A.}\ \bibnamefont {Li}},
  \bibinfo {author} {\bibfnamefont {R.~E.}\ \bibnamefont {Dunin-Borkowski}},
  \bibinfo {author} {\bibfnamefont {R.~W.}\ \bibnamefont {Boyd}}, \bibinfo
  {author} {\bibfnamefont {B.~J.}\ \bibnamefont {McMorran}}, \ and\ \bibinfo
  {author} {\bibfnamefont {E.}~\bibnamefont {Karimi}},\ }\href@noop {}
  {\bibfield  {journal} {\bibinfo  {journal} {Nature Communications}\ }\textbf
  {\bibinfo {volume} {8}},\ \bibinfo {pages} {689} (\bibinfo {year}
  {2017}{\natexlab{b}})}\BibitemShut {NoStop}%
\bibitem [{\citenamefont {McMorran}\ \emph {et~al.}(2011)\citenamefont
  {McMorran}, \citenamefont {Agrawal}, \citenamefont {Anderson}, \citenamefont
  {Herzing}, \citenamefont {Lezec}, \citenamefont {McClelland},\ and\
  \citenamefont {Unguris}}]{Mcmorran:2011}%
  \BibitemOpen
  \bibfield  {author} {\bibinfo {author} {\bibfnamefont {B.~J.}\ \bibnamefont
  {McMorran}}, \bibinfo {author} {\bibfnamefont {A.}~\bibnamefont {Agrawal}},
  \bibinfo {author} {\bibfnamefont {I.~M.}\ \bibnamefont {Anderson}}, \bibinfo
  {author} {\bibfnamefont {A.~A.}\ \bibnamefont {Herzing}}, \bibinfo {author}
  {\bibfnamefont {H.~J.}\ \bibnamefont {Lezec}}, \bibinfo {author}
  {\bibfnamefont {J.~J.}\ \bibnamefont {McClelland}}, \ and\ \bibinfo {author}
  {\bibfnamefont {J.}~\bibnamefont {Unguris}},\ }\href@noop {} {\bibfield
  {journal} {\bibinfo  {journal} {Science}\ }\textbf {\bibinfo {volume}
  {331}},\ \bibinfo {pages} {192} (\bibinfo {year} {2011})}\BibitemShut
  {NoStop}%
\bibitem [{\citenamefont {Grillo}\ \emph {et~al.}(2014)\citenamefont {Grillo},
  \citenamefont {Carlo~Gazzadi}, \citenamefont {Karimi}, \citenamefont
  {Mafakheri}, \citenamefont {Boyd},\ and\ \citenamefont
  {Frabboni}}]{Grillo:2014}%
  \BibitemOpen
  \bibfield  {author} {\bibinfo {author} {\bibfnamefont {V.}~\bibnamefont
  {Grillo}}, \bibinfo {author} {\bibfnamefont {G.}~\bibnamefont
  {Carlo~Gazzadi}}, \bibinfo {author} {\bibfnamefont {E.}~\bibnamefont
  {Karimi}}, \bibinfo {author} {\bibfnamefont {E.}~\bibnamefont {Mafakheri}},
  \bibinfo {author} {\bibfnamefont {R.~W.}\ \bibnamefont {Boyd}}, \ and\
  \bibinfo {author} {\bibfnamefont {S.}~\bibnamefont {Frabboni}},\ }\href@noop
  {} {\bibfield  {journal} {\bibinfo  {journal} {Applied Physics Letters}\
  }\textbf {\bibinfo {volume} {104}},\ \bibinfo {pages} {043109} (\bibinfo
  {year} {2014})}\BibitemShut {NoStop}%
\bibitem [{\citenamefont {Uchida}\ and\ \citenamefont
  {Tonomura}(2010)}]{Uchida:2010}%
  \BibitemOpen
  \bibfield  {author} {\bibinfo {author} {\bibfnamefont {M.}~\bibnamefont
  {Uchida}}\ and\ \bibinfo {author} {\bibfnamefont {A.}~\bibnamefont
  {Tonomura}},\ }\href@noop {} {\bibfield  {journal} {\bibinfo  {journal}
  {Nature}\ }\textbf {\bibinfo {volume} {464}},\ \bibinfo {pages} {737}
  (\bibinfo {year} {2010})}\BibitemShut {NoStop}%
\bibitem [{\citenamefont {Shiloh}\ \emph {et~al.}(2014)\citenamefont {Shiloh},
  \citenamefont {Lereah}, \citenamefont {Lilach},\ and\ \citenamefont
  {Arie}}]{Shiloh:2014}%
  \BibitemOpen
  \bibfield  {author} {\bibinfo {author} {\bibfnamefont {R.}~\bibnamefont
  {Shiloh}}, \bibinfo {author} {\bibfnamefont {Y.}~\bibnamefont {Lereah}},
  \bibinfo {author} {\bibfnamefont {Y.}~\bibnamefont {Lilach}}, \ and\ \bibinfo
  {author} {\bibfnamefont {A.}~\bibnamefont {Arie}},\ }\href@noop {} {\bibfield
   {journal} {\bibinfo  {journal} {Ultramicroscopy}\ }\textbf {\bibinfo
  {volume} {144}},\ \bibinfo {pages} {26} (\bibinfo {year} {2014})}\BibitemShut
  {NoStop}%
\bibitem [{\citenamefont {Jesacher}\ \emph {et~al.}(2005)\citenamefont
  {Jesacher}, \citenamefont {F{\"u}rhapter}, \citenamefont {Bernet},\ and\
  \citenamefont {Ritsch-Marte}}]{Jesacher:2005}%
  \BibitemOpen
  \bibfield  {author} {\bibinfo {author} {\bibfnamefont {A.}~\bibnamefont
  {Jesacher}}, \bibinfo {author} {\bibfnamefont {S.}~\bibnamefont
  {F{\"u}rhapter}}, \bibinfo {author} {\bibfnamefont {S.}~\bibnamefont
  {Bernet}}, \ and\ \bibinfo {author} {\bibfnamefont {M.}~\bibnamefont
  {Ritsch-Marte}},\ }\href@noop {} {\bibfield  {journal} {\bibinfo  {journal}
  {Physical Review Letters}\ }\textbf {\bibinfo {volume} {94}},\ \bibinfo
  {pages} {233902} (\bibinfo {year} {2005})}\BibitemShut {NoStop}%
\bibitem [{\citenamefont {F{\"u}rhapter}\ \emph {et~al.}(2007)\citenamefont
  {F{\"u}rhapter}, \citenamefont {Jesacher}, \citenamefont {Maurer},
  \citenamefont {Bernet},\ and\ \citenamefont {Ritsch-Marte}}]{Furhapter:2007}%
  \BibitemOpen
  \bibfield  {author} {\bibinfo {author} {\bibfnamefont {S.}~\bibnamefont
  {F{\"u}rhapter}}, \bibinfo {author} {\bibfnamefont {A.}~\bibnamefont
  {Jesacher}}, \bibinfo {author} {\bibfnamefont {C.}~\bibnamefont {Maurer}},
  \bibinfo {author} {\bibfnamefont {S.}~\bibnamefont {Bernet}}, \ and\ \bibinfo
  {author} {\bibfnamefont {M.}~\bibnamefont {Ritsch-Marte}},\ }\href@noop {}
  {\bibfield  {journal} {\bibinfo  {journal} {Advances in imaging and electron
  physics}\ }\textbf {\bibinfo {volume} {146}},\ \bibinfo {pages} {1} (\bibinfo
  {year} {2007})}\BibitemShut {NoStop}%
\bibitem [{\citenamefont {Karlovets}(2018)}]{Karlovets:2018}%
  \BibitemOpen
  \bibfield  {author} {\bibinfo {author} {\bibfnamefont {D.}~\bibnamefont
  {Karlovets}},\ }\href@noop {} {\bibfield  {journal} {\bibinfo  {journal}
  {arXiv preprint arXiv:1803.09150}\ } (\bibinfo {year} {2018})}\BibitemShut
  {NoStop}%
\bibitem [{\citenamefont {Pozzi}\ \emph {et~al.}(2017)\citenamefont {Pozzi},
  \citenamefont {Lu}, \citenamefont {Tavabi}, \citenamefont {Duchamp},\ and\
  \citenamefont {Dunin-Borkowski}}]{Pozzi:2017}%
  \BibitemOpen
  \bibfield  {author} {\bibinfo {author} {\bibfnamefont {G.}~\bibnamefont
  {Pozzi}}, \bibinfo {author} {\bibfnamefont {P.-H.}\ \bibnamefont {Lu}},
  \bibinfo {author} {\bibfnamefont {A.~H.}\ \bibnamefont {Tavabi}}, \bibinfo
  {author} {\bibfnamefont {M.}~\bibnamefont {Duchamp}}, \ and\ \bibinfo
  {author} {\bibfnamefont {R.~E.}\ \bibnamefont {Dunin-Borkowski}},\
  }\href@noop {} {\bibfield  {journal} {\bibinfo  {journal} {Ultramicroscopy}\
  }\textbf {\bibinfo {volume} {181}},\ \bibinfo {pages} {191} (\bibinfo {year}
  {2017})}\BibitemShut {NoStop}%
\bibitem [{\citenamefont {Verbeeck}\ \emph {et~al.}(2018)\citenamefont
  {Verbeeck}, \citenamefont {B{\'e}ch{\'e}}, \citenamefont
  {M{\"u}ller-Caspary}, \citenamefont {Guzzinati}, \citenamefont {Luong},\ and\
  \citenamefont {Den~Hertog}}]{Verbeeck:2018}%
  \BibitemOpen
  \bibfield  {author} {\bibinfo {author} {\bibfnamefont {J.}~\bibnamefont
  {Verbeeck}}, \bibinfo {author} {\bibfnamefont {A.}~\bibnamefont
  {B{\'e}ch{\'e}}}, \bibinfo {author} {\bibfnamefont {K.}~\bibnamefont
  {M{\"u}ller-Caspary}}, \bibinfo {author} {\bibfnamefont {G.}~\bibnamefont
  {Guzzinati}}, \bibinfo {author} {\bibfnamefont {M.~A.}\ \bibnamefont
  {Luong}}, \ and\ \bibinfo {author} {\bibfnamefont {M.}~\bibnamefont
  {Den~Hertog}},\ }\href@noop {} {\bibfield  {journal} {\bibinfo  {journal}
  {Ultramicroscopy}\ }\textbf {\bibinfo {volume} {190}},\ \bibinfo {pages} {58}
  (\bibinfo {year} {2018})}\BibitemShut {NoStop}%
\bibitem [{\citenamefont {Aharonov}\ and\ \citenamefont
  {Bohm}(1959)}]{Aharonov:1959}%
  \BibitemOpen
  \bibfield  {author} {\bibinfo {author} {\bibfnamefont {Y.}~\bibnamefont
  {Aharonov}}\ and\ \bibinfo {author} {\bibfnamefont {D.}~\bibnamefont
  {Bohm}},\ }\href@noop {} {\bibfield  {journal} {\bibinfo  {journal} {Physical
  Review}\ }\textbf {\bibinfo {volume} {115}},\ \bibinfo {pages} {485}
  (\bibinfo {year} {1959})}\BibitemShut {NoStop}%
\bibitem [{\citenamefont {Boyer}(1973)}]{Boyer:1973}%
  \BibitemOpen
  \bibfield  {author} {\bibinfo {author} {\bibfnamefont {T.~H.}\ \bibnamefont
  {Boyer}},\ }\href@noop {} {\bibfield  {journal} {\bibinfo  {journal}
  {Physical Review D}\ }\textbf {\bibinfo {volume} {8}},\ \bibinfo {pages}
  {1679} (\bibinfo {year} {1973})}\BibitemShut {NoStop}%
\bibitem [{\citenamefont {Blackburn}(2016)}]{Blackburn:2016}%
  \BibitemOpen
  \bibfield  {author} {\bibinfo {author} {\bibfnamefont {A.~M.}\ \bibnamefont
  {Blackburn}},\ }\href@noop {} {\bibfield  {journal} {\bibinfo  {journal}
  {Microscopy and Microanalysis}\ }\textbf {\bibinfo {volume} {22}},\ \bibinfo
  {pages} {1710} (\bibinfo {year} {2016})}\BibitemShut {NoStop}%
\end{thebibliography}
\end{document}